\documentclass[%
 reprint,
superscriptaddress,
 amsmath,amssymb,
 aps,
]{revtex4-2}

\usepackage{placeins}

\usepackage{graphicx}
\usepackage{dcolumn}
\usepackage{bm}
\usepackage{braket}
\usepackage{color}
\usepackage{amsthm}
\usepackage{dsfont}
\usepackage[absolute]{textpos}
\usepackage{xcolor}
\usepackage{fancyhdr}
\usepackage{listings}
\usepackage{algorithmic}
\usepackage[ruled,vlined]{algorithm2e}
\usepackage{listings}
\usepackage{xcolor}
\usepackage{url}
\definecolor{darkblue}{rgb}{0,0,0.5}
\usepackage{hyperref}
\hypersetup{
colorlinks=true,
linkcolor=black,
filecolor=blue,
citecolor=darkblue,  
urlcolor=black,
}

\definecolor{codegreen}{rgb}{0,0.6,0}
\definecolor{codegray}{rgb}{0.5,0.5,0.5}
\definecolor{codepurple}{rgb}{0.58,0,0.82}
\definecolor{backcolour}{rgb}{0.95,0.95,0.92}

\lstdefinestyle{mystyle}{
    backgroundcolor=\color{backcolour},   
    commentstyle=\color{codegreen},
    keywordstyle=\color{magenta},
    numberstyle=\tiny\color{codegray},
    stringstyle=\color{codepurple},
    basicstyle=\ttfamily\footnotesize,
    breakatwhitespace=false,         
    breaklines=true,                 
    captionpos=b,                    
    keepspaces=true,                 
    numbers=left,                    
    numbersep=5pt,                  
    showspaces=false,                
    showstringspaces=false,
    showtabs=false,                  
    tabsize=2
}

\DeclareMathOperator{\Tr}{Tr}

\def\bx{\mathbf{x}}
\def\by{\mathbf{y}}
\def\bz{\mathbf{z}}
\def\bw{\mathbf{w}}

\def\balpha{\bm{\alpha}}

\def\bbD{{\bm{\mathcal{D}}}}

\usepackage{appendix}

\preprint{APS/123-QED}

\begin{document}
\title{Demonstration of machine-learning-enhanced Bayesian quantum state estimation}

\author{Sanjaya Lohani}
\email[]{slohan3@uic.edu}
\affiliation{Dept.~of Electrical \& Computer Engineering, University of Illinois Chicago, Chicago, IL 60607, USA}
\author{Joseph M. Lukens}
\affiliation{Research Technology Office, Arizona State University, Tempe, Arizona 85287, USA}
\affiliation{Quantum Information Science Section, Oak Ridge National Laboratory, Oak Ridge, TN 37831, USA}
\author{Atiyya A. Davis}
\affiliation{Dept.~of Electrical \& Computer Engineering, University of Illinois Chicago, Chicago, IL 60607, USA}
\author{Amirali Khannejad}
\affiliation{Dept.~of Electrical \& Computer Engineering, University of Illinois Chicago, Chicago, IL 60607, USA}
\author{Sangita Regmi}
\affiliation{Dept.~of Electrical \& Computer Engineering, University of Illinois Chicago, Chicago, IL 60607, USA}
\author{Daniel E. Jones}
\affiliation{DEVCOM US Army Research Laboratory, Adelphi, MD 20783, USA}
\author{Ryan~T. Glasser}
\affiliation{Tulane University, New Orleans, LA 70118, USA}
\author{Thomas~A. Searles}
\email[]{tsearles@uic.edu}
\affiliation{Dept.~of Electrical \& Computer Engineering, University of Illinois Chicago, Chicago, IL 60607, USA}
\author{Brian~T. Kirby}
\email[]{brian.t.kirby4.civ@army.mil}
\affiliation{DEVCOM US Army Research Laboratory, Adelphi, MD 20783, USA}
\affiliation{Tulane University, New Orleans, LA 70118, USA}


\begin{abstract}
Machine learning (ML) has found broad applicability in quantum information science in topics as diverse as experimental design, state classification, and even studies on quantum foundations. Here, we experimentally realize an approach for defining custom prior distributions that are automatically tuned using ML for use with Bayesian quantum state estimation methods. Previously, researchers have looked to Bayesian quantum state tomography due to its unique advantages like natural uncertainty quantification, the return of reliable estimates under any measurement condition, and minimal mean-squared error. However, practical challenges related to long computation times and conceptual issues concerning how to incorporate prior knowledge most suitably can overshadow these benefits. Using both simulated and experimental measurement results, we demonstrate that ML-defined prior distributions reduce net convergence times and provide a natural way to incorporate both implicit and explicit information directly into the prior distribution. These results constitute a promising path toward practical implementations of Bayesian quantum state tomography. 
\end{abstract}
\maketitle

\section{Introduction}

Quantum state tomography (QST) is a process for characterizing the state of a quantum system through a sequence of measurements. 
In general, QST consists of two distinct steps: the repeated measurement of identically prepared systems 
and the entirely classical post-processing step of determining the density matrix most consistent with the measurement results. However, given the substantial experimental challenges associated with collecting measurements of high-dimensional quantum systems, the computational resources involved with state reconstruction have only relatively recently become a significant bottleneck in their own right~\cite{hou_full_2016, smolin_efficient_2012}. Various approaches for performing state estimation given a set of measurement results appear in the literature, including maximum likelihood \cite{Hradil1997, James2001, james2005measurement, teo2011quantum, smolin_efficient_2012}, compressive sensing \cite{Gross2010, liu2012experimental}, machine learning (ML) \cite{carrasquilla2019reconstructing, cha2020attention, tiunov2020experimental, torlai2019integrating, neugebauer2020neural, lohani2020machine, xu2018neural, Lohani_2022,ahmed2021quantum}, 
and Bayesian inference \cite{Blume2010, Seah2015, Granade2016, Williams2017,Mai2017, lukens2020practical,lukens2021bayesian}. The existing inference methods define a trade-space between different characteristics such as general applicability and assumptions about the system, computation time, and accuracy guarantees. 

Among the available state estimation approaches, Bayesian estimation is the only procedure that naturally includes uncertainty quantification, the return of reliable estimates under any measurement condition, and minimal mean-squared error \cite{Robert1999}. Unfortunately, these advantages come with the practical computational challenges related to the calculation of high-dimensional integrals, although the estimation of these integrals through Monte Carlo methods has proven useful~\cite{Blume2010, Seah2015, Granade2016,Williams2017,Mai2017,lukens2020practical,lukens2021bayesian}. 
Further, efforts to include prior information in the Bayesian estimation framework are often ad hoc, relying on custom likelihood functions or the direct manipulation of prior distributions. 
While the former has shown some promise in terms of improving efficiency \cite{Mai2017, lukens2020practical}, it is conceptually unsatisfying, 
as in principle all prior information should enter through the prior distribution. On the other hand, current approaches for including prior information through direct engineering of prior distributions are coarse and rely on assumptions about a system that may change over time, such as the expected rank of output states, while also requiring manual tuning \cite{lohani2021improving,Lohani_2022}.

\begin{figure}[ht!]
\centering
\includegraphics[width=\linewidth]{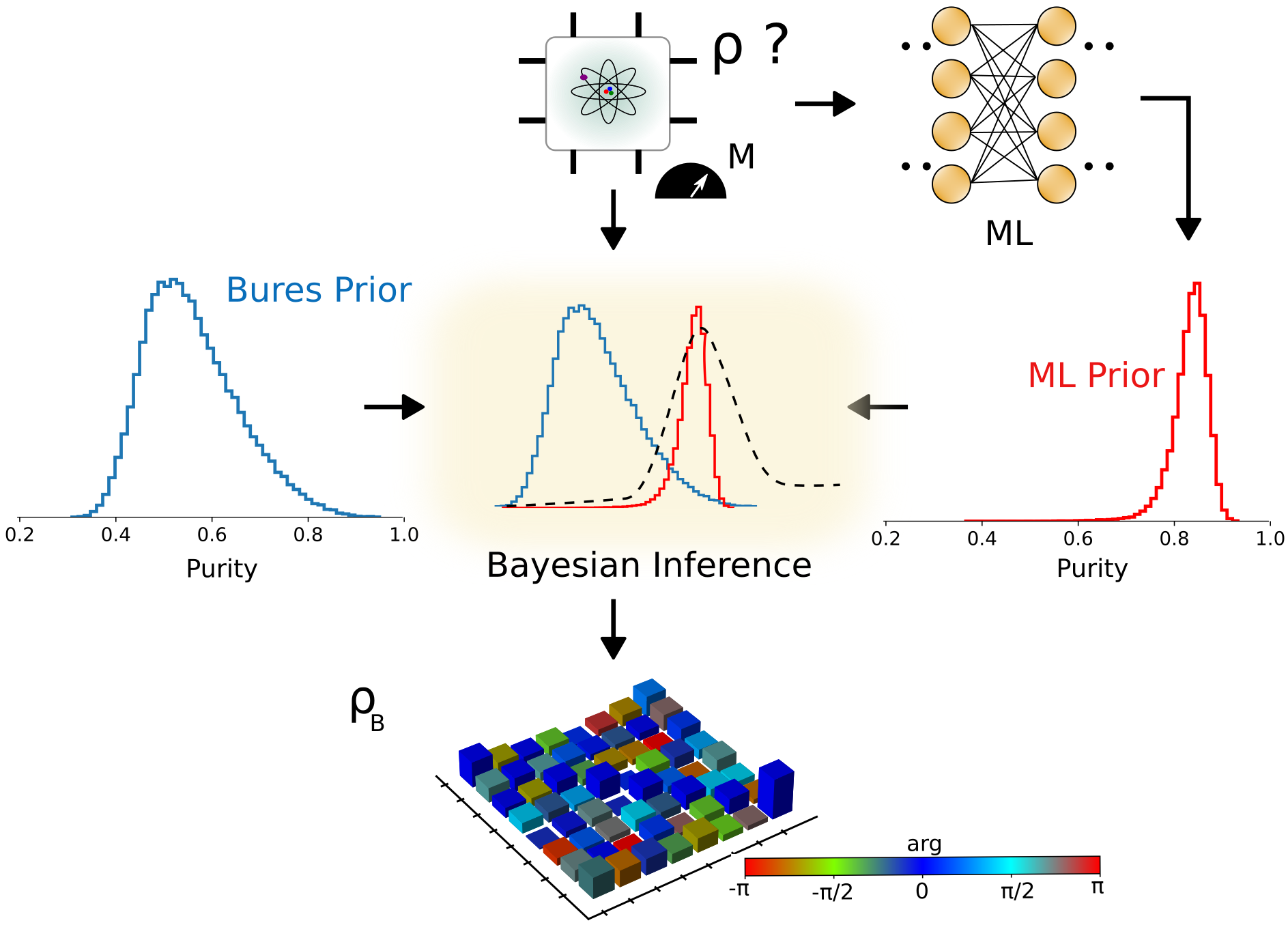}
\caption{Outline of proposed ML-prior Bayesian estimation method. 
Shown are the measurement results entering (top of figure) and two different potential prior distributions: a fixed prior in blue (Bures, in this example) and the ML prior, which adapts to measurement results (shown in red). The dashed line represents the actual probability density function of the experimental scenario and emphasizes that the Bures prior is unchanged  while the ML-prior distribution adapts by incorporating information from the measurement results. The output of the system, given either prior distribution, is the Bayesian mean density matrix as shown at the bottom of the figure.} 
\label{fig:intro}
\end{figure}

Here we propose and experimentally realize a method, pictured in Fig. \ref{fig:intro}, for mitigating Bayesian state reconstruction's practical and conceptual challenges through ML-defined prior distributions that automatically adapt to input data. 
In particular, we use a pre-trained neural network reconstruction system to define a custom prior distribution for any arbitrary input dataset and to reconstruct the state of an experimental system consisting of two entangled polarization qubits. 
Our approach is sufficiently general to encompass several previous methods for manually tuned prior distributions and allows for either direct or automatic tuning of prior distribution parameters. We find a net reduction in the time required for Bayesian inference to reach a given reconstruction fidelity, even when including the additional time required by the pre-trained neural network to define the custom prior distribution. We first demonstrate the efficacy of our approach on simulated measurement results for systems of up to four qubits and then experimentally with a fiber-based source of polarization-entangled photons. 
In all cases, our ML-enhanced Bayesian inference approach reaches a lower-error estimate with reduced computational resources than a traditional prior distribution. Moreover, this advantage persists even when the traditional prior for comparison is the same as that used to generate the test set, the precise scenario where Bayesian quantum state reconstruction is guaranteed to minimize mean squared error.

\section{Bayesian quantum state reconstruction}\label{Bayesian state reconstruction}


Intuitively, reconstruction of quantum states using Bayesian inference takes some prior distribution (an initial set of beliefs regarding what states might describe the system, possibly based on prior information, or intentionally chosen to be uniform) and, through Bayes' rule, incorporates measurements to update the prior and generate a posterior distribution. Hence, unlike other approaches to quantum state reconstruction, such as maximum likelihood~\cite{Hradil1997, James2001}, Bayesian inference produces a complete distribution of possible states whose average minimizes mean squared error~\cite{Robert1999}. In order to perform this Bayesian updating, we first require a parameterized description of a density matrix, given by $\rho(\bw)$, where all possible values of the vector of parameters $\bw$ result in a valid physical density matrix---i.e., unit-trace, Hermitian, and positive-semidefinite. Further, we must define a prior distribution $\pi_{0}(\bw)$ over these parameters, possibly containing descriptive information.

Then, following an experiment, Bayes' rule defines the posterior distribution for $\bw$ as
\begin{equation}
\label{j1}
\pi(\bw) = \frac{1}{\mathcal{Z}} L_\bbD(\bw) \pi_0(\bw),
\end{equation}
where the likelihood $L_\bbD(\bw)\propto P(\bbD|\bw)$ (the probability of observing dataset $\bbD$ given parameters $\bw$) and $\mathcal{Z}$ is a normalizing constant (which in practice often does not need to be determined explicitly). 
In the context of quantum state reconstruction the likelihood can be expressed as $L_\bbD(\bw) = \prod_{m=1}^M \Tr \rho(\bw) \Lambda_m$ where each of $M$ measurements, described by an operator $\Lambda_m$, is performed on identically prepared input states. As QST in practical settings is often based on the simpler class of projective measurements, we can further express the likelihood as
\begin{equation}
\label{j2}
L_\bbD(\bw) = \prod_{m=1}^M \braket{\psi_m|\rho(\bw)|\psi_m},
\end{equation}
with $\ket{\psi_m}$ the eigenstate observed in measurement $m$.
The Bayesian mean estimate of the reconstructed quantum density matrix $\rho$ follows as
\begin{equation}
\label{j3}
\rho_{B} = \int d\bw\, \pi(\bw) \rho(\bw).
\end{equation}
The computation of this integral is frequently a severe roadblock to implementation. While several methods for estimating this integral exist, we utilize preconditioned Crank--Nicolson (pCN) Markov chain Monte Carlo (MCMC) in this investigation~\cite{Cotter2013, lukens2020practical}.

Regardless of the specific prior distribution, as long as it provides nonzero support to the full Hilbert space and sufficient measurement results are available, Bayes' rule will converge to a sharply peaked single answer. However, in several important contexts, the selection of a prior distribution can significantly impact the performance of a Bayesian reconstruction system, for example 
whenever experimental data are severely limited or incomplete. 
In addition, even when unlimited data are available, typically the further the prior distribution deviates from the ground truth, the longer it will take computational methods to sample the posterior. 

Previous work considering the selection of prior distributions for Bayesian reconstruction has often defaulted to standard distributions based on fair sampling states according to the Bures~\cite{sommers2003bures} or Hilbert--Schmidt~\cite{zyczkowski2003hilbert} measure. Unfortunately, these distributions are, on average, significantly more mixed than many relevant experimental scenarios and hence have motivated the use of alternatives such as the Mai--Alquier (MA) distribution~\cite{Mai2017}, which can be tuned to favor lower-rank states than those obtained on average from Bures or Hilbert--Schmidt metrics~\cite{lohani2021improving, Lohani_2022}. Recent results show that manually tuning the MA distribution with even very coarse knowledge, such as the mean state purity expected in an experimental scenario, can improve the performance of Bayesian reconstruction methods~\cite{lohani2021improving}. Unfortunately, these approaches require manually updating the prior distribution if experimental conditions change. 

\section{Bayesian inference with an ML-defined prior}\label{ml-prior}



Significant recent research effort has focused on leveraging ML to perform quantum state reconstruction. While different results reveal that ML-based reconstruction potentially has advantages over conventional approaches when reconstructing data that include experimental errors, missing measurements, or high statistical noise, the primary motivator has been to reduce computation times~\cite{Lohani_2022}. In particular, ML-based reconstruction methods that utilize pre-trained networks effectively ``frontload'' expensive computations into training the network so that, once trained, it can perform inference indefinitely without further computationally intensive training~\cite{lohani2021experimental}. However, despite these potential advantages, ML-based reconstruction does not provide absolute performance guarantees, error quantification, or a natural way to incorporate prior information without further training.

Here we propose leveraging the main advantage of ML-based quantum state reconstruction---rapid inference---to automatically define prior distributions for use with Bayesian inference. In this way, we can translate some of the reduced computation time from pre-trained networks to lower the net time required for Bayesian inference. In particular, we propose using a pre-trained ML-based reconstruction system to first supply a rapid estimate $\rho_{ML}$ based on input measurement data, which in turn serves to bias the prior distribution. Then, similar to the MA distribution~\cite{Mai2017}, we generate the complete prior distribution through a Dirichlet-weighted convex sum of $\rho_{ML}$ and Haar-distributed random pure states.

The Dirichlet distribution is defined for vectors $\mathbf{x}=(x_{1},...,x_{K})$ whose elements belong to the open $K-1$ simplex. 
The probability density function of the Dirichlet distribution is defined by
\begin{equation}
    \text{Dir}(\mathbf{x}\vert\bm{\alpha})=\frac{\Gamma\left(\sum_{i=1}^{K}\alpha_{i}\right)}{\prod_{i=1}^{K}\Gamma(\alpha_{i})}\prod_{i=1}^{K}x_{i}^{\alpha_{i}-1},
    \label{eq:dirichlet}
\end{equation}
where $\bm{\alpha}=(\alpha_{1},...,\alpha_{K})$ with all $\alpha_{i}\ge 0$ defines the concentration parameters and $\Gamma(\cdot)$ denotes the standard gamma function. As we will see in a moment, $\bm{\alpha}$ are free parameters that determine the moments of the distribution.

One special case results for $\alpha_{1}=\alpha_{a}$ and $\alpha_{2\leq i\leq K}=\alpha_{b}$.
Then the sum of the $\alpha_{i}$ becomes $\alpha_{0}=\alpha_{a}+(K-1)\alpha_{b}$, and the first-order moments are
\begin{equation}
\begin{aligned}
    \text{E}(X_{1})&=\frac{\alpha_{a}}{\alpha_0}\\
    \text{E}(X_{2\leq i\leq K})&=\frac{\alpha_{b}}{\alpha_{0}}.\\
\end{aligned}
\label{eq:biased_average_expectation}
\end{equation}
We can subsequently define a family of distributions of $D$-dimensional quantum states consisting of the convex sum of $\rho_{ML}$ and $K-1$ Haar-random pure states $\ket{\psi_{i}}$ as
\begin{equation}
\rho=x_{1}\rho_{ML}+\sum_{i=2}^{K}x_{i}\vert\psi_{i}\rangle\langle\psi_{i}\vert.
    \label{eq:bias}
\end{equation}
The degrees of freedom of this family of distributions are the system dimension $D$, the number of terms in the convex sum $K$, and the two concentration parameters $\alpha_{a}$ and $\alpha_{b}$. 
For simplicity, and motivated by previous related work using similar distributions of quantum states \cite{lohani2021improving}, we will set $K=D+1$.
The choice of $K=D+1$ means the sum of Haar-random pure states in $\rho$ has $D$ terms and hence, regardless of $\rho_{ML}$, $\rho$ is full-rank with high probability.
Full-rank estimates are desirable in this context as we want to construct prior distributions that are capable of representing any state of a $D$-dimensional Hilbert space. 

As a particularly convenient reparametrization, we consider
the sum $\alpha_0$ above and define the ratio $\mu$ as
\begin{equation}
   \mu= \frac{\alpha_{a}}{\alpha_{b}} = \frac{\text{E}(X_{1})}{\text{E}(X_{2\leq i\leq K})},
\end{equation}
so that the expectation values become
\begin{equation}
\begin{aligned}
    \text{E}(X_{1})&=\frac{\mu}{K+\mu-1}\\
    \text{E}(X_{2\leq i\leq K})&=\frac{1}{K+\mu-1}\\
\end{aligned}
\label{eq:mean_mu}
\end{equation}
and variances
\begin{equation}
\begin{aligned}
\text{var}(X_{1})&=\frac{\mu  (K-1)}{(\alpha_{0}+1) (\mu +K-1)^2}\\
\text{var}(X_{2\leq i\leq K})&=\frac{\mu +K-2}{(\alpha_{0}+1) (\mu +K-1)^2}.
\end{aligned}
\end{equation}
Thus, the mean values are determined only by $\mu$, while both $\mu$ and $\alpha_{0}$ contribute to the variances. 

In particular, $\alpha_{0}$ determines the average sparsity of the Dirichlet vector $\mathbf{x}$ while $\mu$ determines the bias of the distribution toward $\rho_{ML}$. 
The sparsity effects of $\alpha_{0}$ are most readily apparent when $\mu=1$, meaning the Dirichlet parameters are unbiased. Here a flat distribution of the entire $K$-simplex is obtained when $\alpha_{0}=1$; a tendency for all components of $\mathbf{x}$ to equal $1/K$ occurs when $\alpha_{0}\gg 1$; and finally, the distribution becomes sparse as $\alpha_{0}\rightarrow 0$, with one component equal to unity (at random) and all others zero.
Alternatively, to see the impact of altering $\mu$ we can express the mean of the ML-prior distribution as $\text{E}(\rho) \propto \mu\rho_{ML} + \frac{(K-1)}{D}I_D$, with $I_D$ being the $D\times D$ identity and where we have used the expectation values from Eq.~(\ref{eq:mean_mu}).
Unlike a uniform prior that averages to the maximally mixed state $\frac{1}{D}I_D$, this state is biased towards $\rho_{ML}$ with a weight determined by $\mu$. 
For example, for a fixed $\alpha_0$, as $\mu\rightarrow\infty$ the distribution approaches a sharp peak around $\rho_{ML}$; the opposite limit $\mu\rightarrow 0$ returns a uniform prior with $\text{E}(\rho) = \frac{1}{D} I_{D}$.

Interestingly, the state construction defined by Eq.~(\ref{eq:bias}) with Dirichlet-distributed coefficients [Eq.~(\ref{eq:dirichlet})] is capable of incorporating both implicit and explicit prior information. The information contained within $\rho_{ML}$ can be viewed as \emph{implicit} since $\rho_{ML}$ is automatically generated based on the measurement inputs and hence requires no prior information about the experimental situation.
Alternatively, we can manually tune $\mu$ and $\alpha_{0}$ to reflect \emph{explicit} prior information. As described above these two parameters allow us to control both the sparsity (related to the mean purity of the distribution) and bias towards $\rho_{ML}$ (related to our confidence in the estimate $\rho_{ML}$).


\begin{figure}[ht]
\centering
\includegraphics[width=\linewidth]{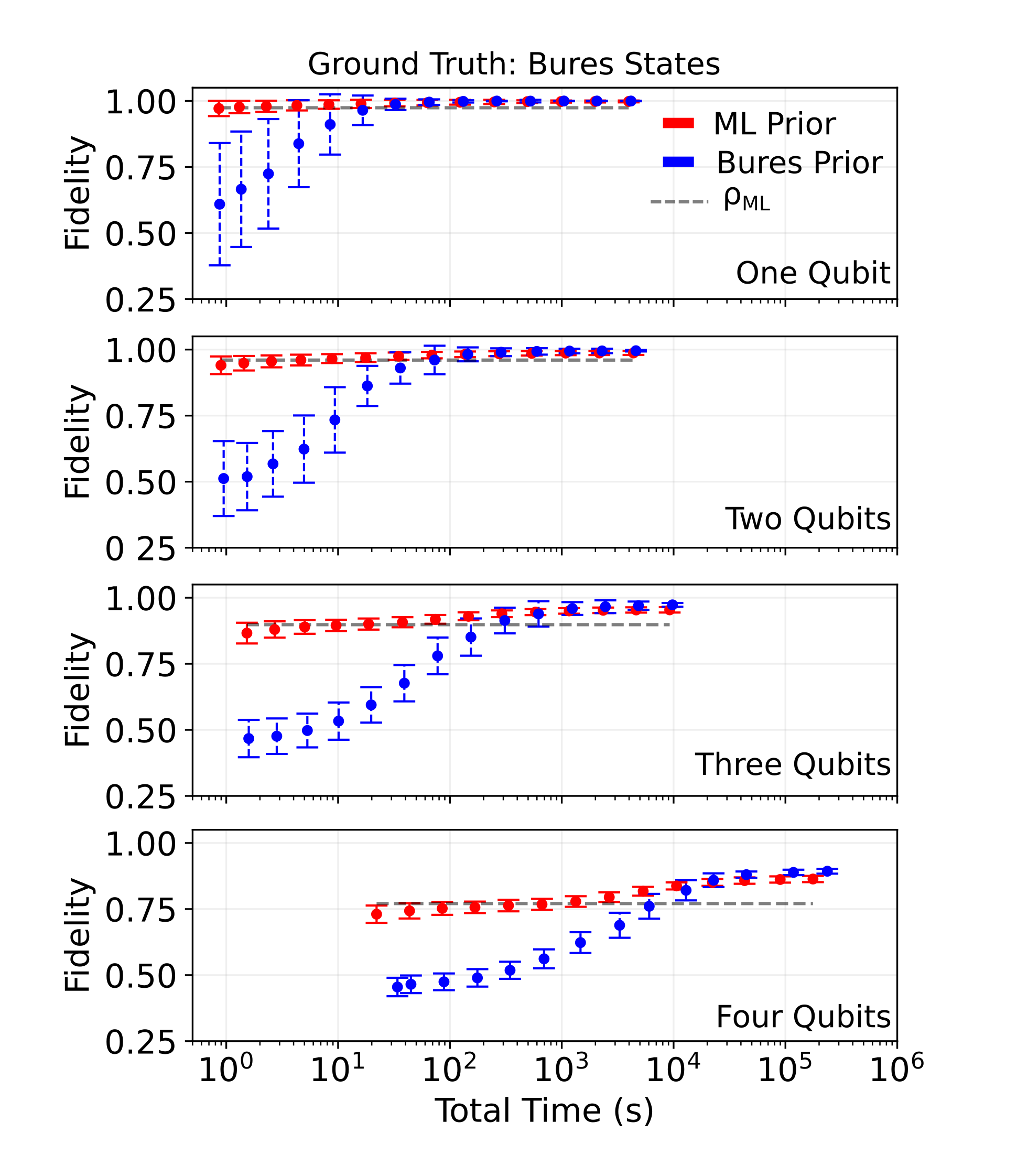}
\caption{Quantum state reconstruction fidelity versus total wall-time 
for Bures-drawn ground-truth states, from one (top) to four (bottom) qubits. Dotted horizontal lines show the average fidelity of ML-reconstructed density matrices for the given test sets. Red dot and blue dot plots represent the cases with ML and Bures priors, respectively. 
Error bars follow one standard deviation from the mean. Note that ML is pre-trained with random states from Bures distribution.} 
\label{fig:theory-test}
\end{figure}

We now test the performance of Bayesian state reconstruction with an ML-defined-prior 
using simulated measurement results for systems of one to four qubits. 
In particular, we consider ground-truth states sampled uniformly according to the Bures measure and
reconstruct the states using either our ML-defined-prior or a more traditional Bures prior. 
In the case of the ML prior, we first train and save an ML model as described in Appendix \ref{machine learning model} with quantum states generated using a fixed distribution (Bures in this section, MA in the next~\cite{Mai2017}). 
Note that the underlying neural network used to generate $\rho_{ML}$ is pre-trained and remains unchanged regardless of experimental inputs; hence any adaptability of the ML prior is a reflection of the inherent flexibility of the ML approach and not a case-by-case incorporation of information.


The reconstruction fidelities with ML-based (with $\mu=25$ and $\alpha_0=11.6$) and Bures priors are shown in Fig.~\ref{fig:theory-test} for successively doubled MCMC chain lengths from $2^5$ to $2^{18}$, where the x-axis includes the time required to compute $\rho_{ML}$.
All times are based on a desktop computer with an 11th Generation Intel i7-11700K processor operating at 3.60~GHz with 16 threads, a GeForce RTX 3080 graphical processing unit (GPU), and 32~GB of RAM. Note that the GPU is only used to make predictions for $\rho_{ML}$.
For a fair comparison, the ML-prior times do omit the neural network training period, since this represents a one-time cost which is not repeated.
The results shown in Fig.~\ref{fig:theory-test}, and the accompanying error bars, are the result of averaging results from $200$ randomly sampled states according to the Bures measure.
In all cases the ML-defined and Bures priors each eventually reach similar reconstruction fidelities, before the Bures prior marginally outperforms the ML one, consistent with 
the known mean-squared optimality of a prior which precisely matches the ground-truth draws~\cite{Robert1999}.
However, the ML-defined prior significantly outperforms Bures at shorter times:
in all four scenarios in Fig.~\ref{fig:theory-test}(a) the net time to reach a low-error estimate is reduced using the ML-defined prior compared to the ideal situation of prior and ground-truth states being sampled from the same distribution.


Intuitively, 
the ML-defined prior biases the Bayesian reconstruction process to ``start'' nearer to the final reconstruction, as highlighted by the horizontal lines in Fig.~\ref{fig:theory-test} marking the mean fidelity of $\rho_{ML}$, the neural-network-produced guess based on the measurement results; the more accurate this initial guess, the better the head start. Ultimately, Bayesian inference with either the ML-based or traditional Bures prior succeeds in surpassing this original ML estimate, but the former is able to do so more rapidly than the latter which
begins further from the final reconstruction. 
However, as we will see in the next section, the benefit of rapid convergence from biasing the distribution toward the initial guess can come at the cost of ultimately limiting the peak reconstruction fidelity. Hence, a tradeoff between improving the short-computation-time reconstruction fidelities and maximizing the long-computation-time fidelities will emerge.



\section{Performance in optical experiment}

\begin{figure}\centering 
	\includegraphics[width=\columnwidth]{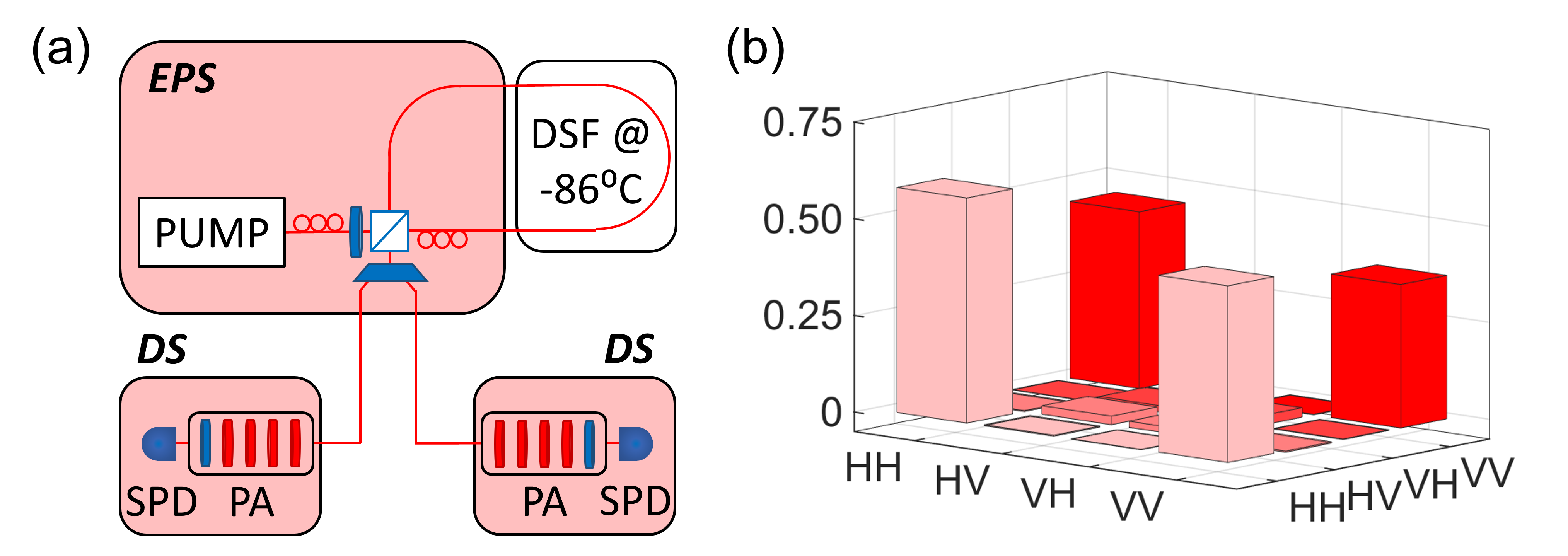}
	\caption{
	(a) Experimental setup used to compare the purity distribution of states output by a polarization-entangled photon source (EPS) with the ML prior and Bures prior purity distributions. DSF: dispersion-shifted fiber. DS: detector station. PA: polarization analyzer consisting of several waveplates (red) and a polarizer (blue). SPD: single-photon detector. 
	(b) Real part of an example density matrix calculated by the ML-prior Bayesian estimation method.}
	\label{epsfig}
\end{figure}

To demonstrate the efficacy of our ML-defined prior approach in a real-world scenario, we apply it to states generated by a commercially available polarization-entangled photon pair source~\cite{nucrypt}. The experimental setup includes an entangled photon source (EPS) connected to two detector stations (DSs) with telecom optical fibers. A schematic of the setup is shown in Fig.~\ref{epsfig}(a), along with the real part of an example density matrix in Fig.~\ref{epsfig}(b), calculated from experimental data using the ML-prior Bayesian estimation method (details below). The EPS outputs photon pairs via four-wave mixing in a dispersion-shifted fiber (DSF) \cite{fiorentino2002all} pumped by a 50~MHz pulsed fiber laser centered at 1552.52~nm. The DSF is cooled to $-86^{\circ}$C in a laboratory-grade freezer to reduce the amount of Raman-scattered noise photons. For the experiment performed here, the average number of photon pairs generated per pulse is set to $\sim$0.1. The signal and idler photons are entangled in polarization by arranging the DSF in a Sagnac loop with a polarizing beam splitter (PBS) and are then spectrally demultiplexed into 100 GHz-spaced ITU outputs~\cite{wang2009robust}.

\begin{figure}[ht]
\centering
\includegraphics[width=.9\linewidth]{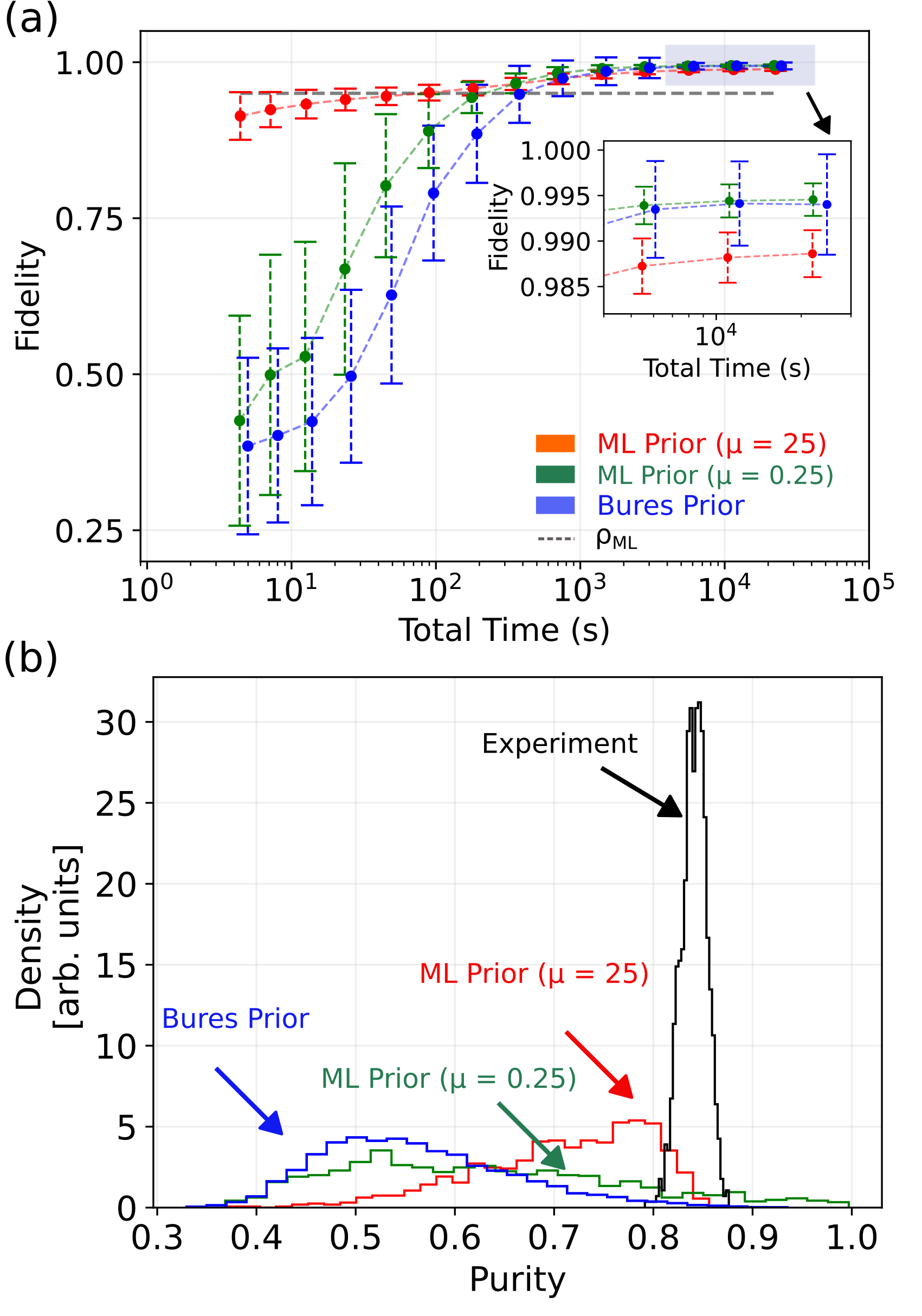}
\caption{Efficacy of ML-prior method for reconstructing states generated by a polarization-entangled photon pair source. (a) Reconstruction fidelity versus total wall-clock time. 
(b) Purity distributions for the ML prior, Bures prior, and experimental states. The error bars represent one standard deviation from the mean. 
}
\label{fig:fid-vs-total-time}
\end{figure}

The detector stations consist of 50~MHz-gated single-photon detectors (SPDs) preceded by polarization analyzers (PAs). The SPDs have detection efficiencies of $\sim$20\% and dark count probabilities of $\sim$4$\times$10$^{-5}$ per gate. Full polarization state tomography is performed by collecting measurements in $36$ different  projections. Maximum coincidence rates of up to several thousand per measurement are achieved by performing each of the 36 measurements over 10 million detector gates. Tomography is performed 1000 consecutive times, resulting in a total experiment duration of $\sim$4~hr. 
The density matrix is calculated for all 1000 runs using Bayesian inference with ML-based and Bures priors. 
As in the previous section, the red curve in Fig.~\ref{fig:fid-vs-total-time}(a) is the reconstruction using the ML prior with $\mu=25$ and $\alpha_0=11.6$, but now utilizing a neural network trained on MA-distributed states with $\alpha=0.4$ to infer $\rho_{ML}$, motivated by previous work that found this distribution balances reconstruction fidelity across the purity spectrum of arbitrary two-qubit states~\cite{lohani2021improving, Lohani_2022}. 
In addition, and again analogous to the previous section, in blue we include the curve showing the reconstruction fidelity assuming a Bures prior.
As before, we see that the ML prior significantly outperforms the Bures prior for short timescales but with the Bures prior ultimately reaching a higher peak fidelity at long computation times.
In some sense, it is surprising that the Bures prior outperforms the ML prior here, considering that---unlike in the previous section---neither distribution matches that of the experimental ground truth. Hence, we would intuitively assume the early lead by the ML prior to persist. One likely explanation for the observed outcome is that the ML prior is in effect \emph{overly} biased toward the initial guess. To test this, we also include the green curve ($\mu=0.25$, $\alpha_{0}=1.7$), for which the ML prior is defined with a significantly reduced bias. This reduced affinity toward $\rho_{ML}$ lowers the initial reconstruction fidelity but, in the limit of long computation times, ultimately outperforms the reconstruction of \emph{both} the previous ML prior (red) and the Bures prior (blue). This interplay highlights the flexibility afforded by our approach via explicit tuning of $\mu$ and $\alpha_0$.

To further understand the behavior in Fig.~\ref{fig:fid-vs-total-time}(a), we also include in Fig.~\ref{fig:fid-vs-total-time}(b) numerically generated probability density functions of purity for the Bures prior (blue), an exemplar of each ML-prior distribution given a randomly chosen set of experimental data (red and green), and the final ML-prior Bayesian-reconstructed density matrices from all 1000 collected experimental datasets (black).
Since the Bures prior is fixed regardless of the experimental scenario we see very little overlap in the probability density functions, as compared with the ML-prior distributions which successfully adapt based on the rapid reconstruction of $\rho_{ML}$. Interestingly, the lower-bias ML prior ($\mu=0.25$) covers the experimental distribution more uniformly than the high-bias case ($\mu=25$), which may contribute to its higher final fidelity in Fig.~\ref{fig:fid-vs-total-time}(a).



\section{Discussion}

In Bayesian inference, the prior distribution ideally encapsulates all prior knowledge about a system. In practice, when performing QST using Bayesian methods, one often selects a prior based on broad experience with a given system and not on a case-by-case basis. This paper has considered a more nuanced approach where a prior distribution is selected after measurement results are collected---and an ML estimate $\rho_{ML}$ obtained---but before Bayesian reconstruction. Ostensibly, such a data-informed prior would seem to complicate the typical Bayesian paradigm that clearly separates initial knowledge (in the prior) from observations (in the likelihood). Nevertheless, our procedure poses no problems mathematically: from the perspective of Bayes' rule, $\rho_{ML}$ is simply a high-dimensional hyperparameter 
whose origin has no impact on the method. Indeed, $\rho_{ML}$ can be viewed more broadly as a target or proposed state, even if it is in fact obtained anachronistically \emph{after} the experiment in question.

It is important to note that the implicit prior defined in this way does not circumvent known optimality conditions for more standard priors. For example, in the tests of Fig.~\ref{fig:theory-test}(a) on Bures-drawn states, Bayesian inference with a Bures prior ultimately converges to higher average fidelities with respect to the ground truth than inference with our ML prior. Such results are consistent with the mean-squared-error optimality of the Bayesian mean estimator when tested on ground-truth states drawn from the prior~\cite{Robert1999}; in other words, irrespective of the process used to tune the prior with ML, the fundamental limit set by a ground-truth-matched prior cannot be surpassed. Of course, in many practical settings the prior is simply a convenient choice and may differ widely from the distribution of states tested, so that adhering strictly to a fiducial distribution like Bures may not be particularly meaningful. In any case, the primary advantages of the ML prior lie in reducing the total computational cost to reach a given inference fidelity, which can be attributed to its initial weighting of the MCMC chain closer to the final converged answer.

Other approaches to Bayesian quantum state reconstruction have also aimed to reduce required computational resources by including dataset-specific information through a pseudo-likelihood based on a simple loss function~\cite{Mai2017,lukens2020practical}, rather than a physics-inspired likelihood like Born's rule in Eq.~(\ref{j2}). While the pseudo-likelihood also leverages low-resolution state reconstructions to improve computational efficiency, the ML prior proposed here is arguably preferable conceptually in that it introduces this information as prior knowledge instead of through modifying the likelihood. Yet regardless of any theoretical distinctions, both approaches offer complementary procedures to utilize some initial estimate 
and improve thereon with Bayesian inference, either through the prior or the likelihood. 

In summary, we have leveraged ML to tune prior distributions in Bayesian quantum state estimation. 
The low computational overhead of our neural network allows us to outperform standard priors in total time to convergence, even when including the computation time required to define the custom prior. Aside from demonstrating this for systems of up to four qubits in simulation, we also apply our approach to two-qubit optical systems and find that the advantages persist. 
By offering a general framework for computational speedups, our work should provide a practical path toward the more widespread adoption of Bayesian inference techniques for quantum state reconstruction.



\section*{Funding} 
U.S. Department of Energy, Office of Science, National Quantum Information Science Research Centers, Co-design Center for Quantum Advantage (C2QA) (DE-SC0012704); Army Research Laboratory and the Army Research Office (W911NF-19-2-0087, W911NF-20-2-0168); U.S. Department of Energy, Office of Science, Advanced Scientific Computing Research (ERKJ353).

\section*{Acknowledgments}
The views and conclusions contained in this document are those of the authors and should not be interpreted as representing the official policies, either expressed or implied, of the Army Research Laboratory or the U.S. Government. The U.S. Government is authorized to reproduce and distribute reprints for Government purposes notwithstanding any copyright notation herein.  A portion of this work was performed at Oak Ridge National Laboratory, operated by UT-Battelle for the U.S. Department of Energy under contract no. DE-AC05-00OR22725.

\smallskip

\section*{Disclosures} The authors declare no conflicts of interest.

\section*{Data Availability Statement} The code and a web-app to generate all datasets are, respectively, openly available at the following URLs: \url{https://github.com/slohani-ai/machine-learning-for-physical-sciences}, and \url{https://mlphys.streamlit.app/}.










\appendix

\section{Machine Learning}\label{machine learning model}
\subsection*{Model}
We design a convolutional neural network (CNN) with a convolutional unit of kernel size (2, 2), strides of 1, ReLU as an activation function, and 25 filters. The output of the CNN is fed into a layer that performs pooling with a pool size (2, 2), followed by a second convolutional unit of the same configuration. Then, we combine two dense layers, followed by a dropout layer with a rate of 0.5, which is then attached to an output layer predicting $\tau$-vectors (Cholesky coefficients of the density matrix {\cite{Banaszek1999, altepeter2005photonic}}). 
We keep the depth (number of hidden layers) of the network fixed for all qubit cases while increasing the number of neurons in the dense layers as shown in Table \ref{table:network layers}.

\begin{table}[h!]
\centering
 \begin{tabular}{|p{1.2cm}|p{1cm}|p{1.3cm}|p{1.3cm}|p{1.3cm}|} 
 \hline
 Qubit Number & Input Shape & Dense Neurons & Dense Neurons & Output Neurons \\ [1ex] 
 \hline\hline
 1 & [2, 3] & 250 & 150 & 4  \\ 
 2 & [6, 6] & 750 & 450 & 16\\
 3 & [6, 36] & 2500 & 1000 & 64 \\
 4 & [36, 36] & 4500 &  2500 & 256 \\ [1ex] 
 \hline
 \end{tabular}
\caption{Number of neurons in fully connected layers as a function of the number of qubits in the state. The $[x,y]$ notation indicates the width ($x$) and length ($y$) of the two-dimensional array of input data after reshaping.}
\label{table:network layers}
\end{table}

\subsection*{Training the model}
To generate the training sets, we sample 40 000 states (for each qubit number) from either the Bures distribution (for Sec.~III of the main text) or the MA distribution with $\alpha=0.4$ 
(for Sec.~IV; see Appendix \ref{MA distribution}), and take 35 000 for training and 5000 for validation. For each state we compute the ground-truth $\tau$ vector and simulate 16 000 random Pauli measurements~\cite{Lohani_2022} (one measurement per each random basis selection), feeding these measurements into the network. The mean square loss between the target and predicted $\tau$ is evaluated and fed back to optimize the network's trainable parameters using the adaptive gradient (Adagrad) optimizer. We use a learning rate of 0.01 and batch size of 100 for up to 75 epochs to train the network.
At the output layer is attached a pipeline that rearranges the predicted $\tau$-vectors into density matrices. The pipeline is built into the same graph of the network for the purposes of computing the average fidelity per epoch for cross-validation and outputting the density matrix directly to avoid post-processing. As an example, in the two-qubit case, the predicted $\tau$-vectors are rearranged to lower triangular matrices $T$ expressed as
\begin{equation}
\begin{aligned}
    &T\,=\,
        \begin{bmatrix}
        \tau_0 & 0& 0& 0\\
        \tau_4+\mathrm{i}\tau_5 & \tau_1 &0 &0\\
        \tau_{10}+\mathrm{i}\tau_{11} & \tau_6+\mathrm{i}\tau_7 &\tau_2 &0\\
        \tau_{14}+\mathrm{i}\tau_{15} & \tau_{12}+\mathrm{i}\tau_{13} &\tau_8+\mathrm{i}\tau_9 &\tau_3\\
        \end{bmatrix}. \\
\end{aligned}
\label{eqn:lower_t}
\end{equation}
The density matrices follow as $\tilde{\rho}=\frac{TT^\dagger}{\textrm{Tr}(TT^\dagger)}$. Note that the physicality of $\tilde{\rho}$ is guaranteed through the Cholesky decomposition, which ensures positive semidefiniteness~\cite{James2001, altepeter2005photonic}. Finally, at the end of the network, the fidelity $F$ between the predicted density matrix $\tilde{\rho}$ and the target $\rho$ is evaluated as $F = \Big|\textrm{Tr}\sqrt{\sqrt{\tilde{\rho}}\rho\sqrt{\tilde{\rho}}}\Big|^2$. An in-depth description of the network architecture is given in \cite{lohani2021experimental}. 

\section{Mai-Alquier (MA) distribution} \label{MA distribution}
This distribution was originally introduced as a prior for Bayesian QST~\cite{Mai2017}, has been applied in several  subsequent studies~\cite{lukens2020practical,lu2020fully,lingaraju2021adaptive,alshowkan2021reconfigurable}, and was recently utilized in \cite{lohani2021improving} to generate training sets for ML-based state reconstruction methods.  The MA distribution is defined as a mixture of Haar-random pure  states with coefficients drawn from the Dirichlet distribution: 
\begin{equation}
    \rho=\sum_{i=1}^{K}x_{i}\vert\psi_{i}\rangle\langle\psi_{i}\vert,
    \label{eq:D_state_def}
\end{equation}
where the vector $\bx$ is a random variable distributed according to $\text{Dir}(\bx|\bm{\alpha})$. This construction matches that of our ML-based prior in Eq.~(6) of the main text, but now with the $i=1$ term a random pure state rather than the fixed $\rho_{ML}$.
If we specialize to the symmetric case $\balpha=\{\alpha,...,\alpha\}$ and assume a Hilbert space dimension of $D$, the expectation value of the purity is
\begin{equation}
        \text{E}_{MA}\left[\text{Tr}(\rho^{2})\right]=\frac{D+\alpha(D+K-1)}{D(1+\alpha K)}.
    \label{eq:purity_av1}
\end{equation}
Finally, we note that in \cite{lohani2021improving} strong evidence was presented that the MA distribution reduces to the HS distribution for $K=\alpha=D$.

\section{Parameterizing the ML prior}\label{param-alpha}

For the ML prior, we take as parameters $\bw=(\by,\bz_2,...,\bz_K)$, where $\by=(y_1,...,y_K)$ is a real $K$-dimensional vector of positive scalars, and $\bz_j$ is a $D$-dimensional complex vector. The prior follows $\pi_0(\bw) \propto y_1^{\alpha_1-1} e^{-y_1}\prod_{i=2}^K y_i^{\alpha_i-1} e^{-y_i} e^{-\frac{1}{2} \bz_i^\dagger \bz_i}$, and we define $\rho(\bw)$ according to
\begin{equation}
\rho(\bw) = \frac{y_1}{\sum_j y_j}\rho_{ML} + \sum_{i=2}^K \left(\frac{y_i}{\sum_j y_j} \right) \frac{\bz_i \bz_i^\dagger}{|\bz_i|^2}.
\label{eq:MLparam}
\end{equation}
As in the main text we take $K=D+1$.
For convenience we utilize independent and unnormalized parameters; our combined gamma and complex-normal prior ensures that the normalized entities $\by/\sum_j y_j$ and $\bz_i/|\bz_i|$ are Dirichlet- and Haar-distributed, respectively, as required~\cite{Mai2017}.

\section{Bures distribution}\label{bures distribution}
A random quantum state $\rho$ from the Bures ensemble can be generated according to
\begin{equation}
        \rho=\frac{\left(\mathds{1}+U\right)G G^{\dagger}\left(\mathds{1}+U^{\dagger}\right)}{\text{Tr}\left[\left(\mathds{1}+U\right)GG^{\dagger}\left(\mathds{1}+U^{\dagger}\right)\right]},
    \label{eq:Bures}
\end{equation}
where $G$ is a random matrix from the Ginibre ensemble and $U$ is a Haar-distributed random unitary from $U(D)$~\cite{al2010random}. Both matrices can be represented as functions of the parameter vector $\bw=(w_1,...,w_{2D^2})$, with each element independently distributed according to a complex standard normal distribution $w_k \sim \mathcal{CN}(0,1)$. We assign $D^2$ of the components to populate the $D\times D$ matrix $G$; the remaining $D^2$ elements comprise a second, independent Ginibre matrix which is then fed into the algorithm of~\cite{Mezzadri2007} to produce the Haar-random unitary $U$. Thus the Bures prior can be written as $\pi_0(\bw) \propto \prod_{k=1}^{2D^2} e^{-\frac{1}{2}|w_k|^2}$.


\bibliographystyle{apsrev4-1}
\bibliography{mlprior}

\begin{thebibliography}{40}%
\makeatletter
\providecommand \@ifxundefined [1]{%
 \@ifx{#1\undefined}
}%
\providecommand \@ifnum [1]{%
 \ifnum #1\expandafter \@firstoftwo
 \else \expandafter \@secondoftwo
 \fi
}%
\providecommand \@ifx [1]{%
 \ifx #1\expandafter \@firstoftwo
 \else \expandafter \@secondoftwo
 \fi
}%
\providecommand \natexlab [1]{#1}%
\providecommand \enquote  [1]{``#1''}%
\providecommand \bibnamefont  [1]{#1}%
\providecommand \bibfnamefont [1]{#1}%
\providecommand \citenamefont [1]{#1}%
\providecommand \href@noop [0]{\@secondoftwo}%
\providecommand \href [0]{\begingroup \@sanitize@url \@href}%
\providecommand \@href[1]{\@@startlink{#1}\@@href}%
\providecommand \@@href[1]{\endgroup#1\@@endlink}%
\providecommand \@sanitize@url [0]{\catcode `\\12\catcode `\$12\catcode
  `\&12\catcode `\#12\catcode `\^12\catcode `\_12\catcode `\%12\relax}%
\providecommand \@@startlink[1]{}%
\providecommand \@@endlink[0]{}%
\providecommand \url  [0]{\begingroup\@sanitize@url \@url }%
\providecommand \@url [1]{\endgroup\@href {#1}{\urlprefix }}%
\providecommand \urlprefix  [0]{URL }%
\providecommand \Eprint [0]{\href }%
\providecommand \doibase [0]{http://dx.doi.org/}%
\providecommand \selectlanguage [0]{\@gobble}%
\providecommand \bibinfo  [0]{\@secondoftwo}%
\providecommand \bibfield  [0]{\@secondoftwo}%
\providecommand \translation [1]{[#1]}%
\providecommand \BibitemOpen [0]{}%
\providecommand \bibitemStop [0]{}%
\providecommand \bibitemNoStop [0]{.\EOS\space}%
\providecommand \EOS [0]{\spacefactor3000\relax}%
\providecommand \BibitemShut  [1]{\csname bibitem#1\endcsname}%
\let\auto@bib@innerbib\@empty
\bibitem [{\citenamefont {Hou}\ \emph {et~al.}(2016)\citenamefont {Hou},
  \citenamefont {Zhong}, \citenamefont {Tian}, \citenamefont {Dong},
  \citenamefont {Qi}, \citenamefont {Li}, \citenamefont {Wang}, \citenamefont
  {Nori}, \citenamefont {Xiang}, \citenamefont {Li},\ and\ \citenamefont
  {Guo}}]{hou_full_2016}%
  \BibitemOpen
  \bibfield  {author} {\bibinfo {author} {\bibfnamefont {Z.}~\bibnamefont
  {Hou}}, \bibinfo {author} {\bibfnamefont {H.-S.}\ \bibnamefont {Zhong}},
  \bibinfo {author} {\bibfnamefont {Y.}~\bibnamefont {Tian}}, \bibinfo {author}
  {\bibfnamefont {D.}~\bibnamefont {Dong}}, \bibinfo {author} {\bibfnamefont
  {B.}~\bibnamefont {Qi}}, \bibinfo {author} {\bibfnamefont {L.}~\bibnamefont
  {Li}}, \bibinfo {author} {\bibfnamefont {Y.}~\bibnamefont {Wang}}, \bibinfo
  {author} {\bibfnamefont {F.}~\bibnamefont {Nori}}, \bibinfo {author}
  {\bibfnamefont {G.-Y.}\ \bibnamefont {Xiang}}, \bibinfo {author}
  {\bibfnamefont {C.-F.}\ \bibnamefont {Li}}, \ and\ \bibinfo {author}
  {\bibfnamefont {G.-C.}\ \bibnamefont {Guo}},\ }\href {\doibase
  10.1088/1367-2630/18/8/083036} {\bibfield  {journal} {\bibinfo  {journal}
  {New J. Phys.}\ }\textbf {\bibinfo {volume} {18}},\ \bibinfo {pages} {083036}
  (\bibinfo {year} {2016})}\BibitemShut {NoStop}%
\bibitem [{\citenamefont {Smolin}\ \emph {et~al.}(2012)\citenamefont {Smolin},
  \citenamefont {Gambetta},\ and\ \citenamefont
  {Smith}}]{smolin_efficient_2012}%
  \BibitemOpen
  \bibfield  {author} {\bibinfo {author} {\bibfnamefont {J.~A.}\ \bibnamefont
  {Smolin}}, \bibinfo {author} {\bibfnamefont {J.~M.}\ \bibnamefont
  {Gambetta}}, \ and\ \bibinfo {author} {\bibfnamefont {G.}~\bibnamefont
  {Smith}},\ }\href {\doibase 10.1103/PhysRevLett.108.070502} {\bibfield
  {journal} {\bibinfo  {journal} {Phys. Rev. Lett.}\ }\textbf {\bibinfo
  {volume} {108}},\ \bibinfo {pages} {070502} (\bibinfo {year}
  {2012})}\BibitemShut {NoStop}%
\bibitem [{\citenamefont {Hradil}(1997)}]{Hradil1997}%
  \BibitemOpen
  \bibfield  {author} {\bibinfo {author} {\bibfnamefont {Z.}~\bibnamefont
  {Hradil}},\ }\href {\doibase 10.1103/PhysRevA.55.R1561} {\bibfield  {journal}
  {\bibinfo  {journal} {Phys. Rev. A}\ }\textbf {\bibinfo {volume} {55}},\
  \bibinfo {pages} {R1561} (\bibinfo {year} {1997})}\BibitemShut {NoStop}%
\bibitem [{\citenamefont {James}\ \emph {et~al.}(2001)\citenamefont {James},
  \citenamefont {Kwiat}, \citenamefont {Munro},\ and\ \citenamefont
  {White}}]{James2001}%
  \BibitemOpen
  \bibfield  {author} {\bibinfo {author} {\bibfnamefont {D.~F.~V.}\
  \bibnamefont {James}}, \bibinfo {author} {\bibfnamefont {P.~G.}\ \bibnamefont
  {Kwiat}}, \bibinfo {author} {\bibfnamefont {W.~J.}\ \bibnamefont {Munro}}, \
  and\ \bibinfo {author} {\bibfnamefont {A.~G.}\ \bibnamefont {White}},\ }\href
  {\doibase 10.1103/PhysRevA.64.052312} {\bibfield  {journal} {\bibinfo
  {journal} {Phys. Rev. A}\ }\textbf {\bibinfo {volume} {64}},\ \bibinfo
  {pages} {052312} (\bibinfo {year} {2001})}\BibitemShut {NoStop}%
\bibitem [{\citenamefont {James}\ \emph {et~al.}(2005)\citenamefont {James},
  \citenamefont {Kwiat}, \citenamefont {Munro},\ and\ \citenamefont
  {White}}]{james2005measurement}%
  \BibitemOpen
  \bibfield  {author} {\bibinfo {author} {\bibfnamefont {D.~F.}\ \bibnamefont
  {James}}, \bibinfo {author} {\bibfnamefont {P.~G.}\ \bibnamefont {Kwiat}},
  \bibinfo {author} {\bibfnamefont {W.~J.}\ \bibnamefont {Munro}}, \ and\
  \bibinfo {author} {\bibfnamefont {A.~G.}\ \bibnamefont {White}},\ }in\
  \href@noop {} {\emph {\bibinfo {booktitle} {Asymptotic Theory of Quantum
  Statistical Inference: Selected Papers}}}\ (\bibinfo  {publisher} {World
  Scientific},\ \bibinfo {year} {2005})\ pp.\ \bibinfo {pages}
  {509--538}\BibitemShut {NoStop}%
\bibitem [{\citenamefont {Teo}\ \emph {et~al.}(2011)\citenamefont {Teo},
  \citenamefont {Zhu}, \citenamefont {Englert}, \citenamefont
  {{\v{R}}eh{\'a}{\v{c}}ek},\ and\ \citenamefont {Hradil}}]{teo2011quantum}%
  \BibitemOpen
  \bibfield  {author} {\bibinfo {author} {\bibfnamefont {Y.~S.}\ \bibnamefont
  {Teo}}, \bibinfo {author} {\bibfnamefont {H.}~\bibnamefont {Zhu}}, \bibinfo
  {author} {\bibfnamefont {B.-G.}\ \bibnamefont {Englert}}, \bibinfo {author}
  {\bibfnamefont {J.}~\bibnamefont {{\v{R}}eh{\'a}{\v{c}}ek}}, \ and\ \bibinfo
  {author} {\bibfnamefont {Z.}~\bibnamefont {Hradil}},\ }\href@noop {}
  {\bibfield  {journal} {\bibinfo  {journal} {Phys. Rev. Lett.}\ }\textbf
  {\bibinfo {volume} {107}},\ \bibinfo {pages} {020404} (\bibinfo {year}
  {2011})}\BibitemShut {NoStop}%
\bibitem [{\citenamefont {Gross}\ \emph {et~al.}(2010)\citenamefont {Gross},
  \citenamefont {Liu}, \citenamefont {Flammia}, \citenamefont {Becker},\ and\
  \citenamefont {Eisert}}]{Gross2010}%
  \BibitemOpen
  \bibfield  {author} {\bibinfo {author} {\bibfnamefont {D.}~\bibnamefont
  {Gross}}, \bibinfo {author} {\bibfnamefont {Y.-K.}\ \bibnamefont {Liu}},
  \bibinfo {author} {\bibfnamefont {S.~T.}\ \bibnamefont {Flammia}}, \bibinfo
  {author} {\bibfnamefont {S.}~\bibnamefont {Becker}}, \ and\ \bibinfo {author}
  {\bibfnamefont {J.}~\bibnamefont {Eisert}},\ }\href {\doibase
  10.1103/PhysRevLett.105.150401} {\bibfield  {journal} {\bibinfo  {journal}
  {Phys. Rev. Lett.}\ }\textbf {\bibinfo {volume} {105}},\ \bibinfo {pages}
  {150401} (\bibinfo {year} {2010})}\BibitemShut {NoStop}%
\bibitem [{\citenamefont {Liu}\ \emph {et~al.}(2012)\citenamefont {Liu},
  \citenamefont {Zhang}, \citenamefont {Liu}, \citenamefont {Chen},\ and\
  \citenamefont {Yuan}}]{liu2012experimental}%
  \BibitemOpen
  \bibfield  {author} {\bibinfo {author} {\bibfnamefont {W.-T.}\ \bibnamefont
  {Liu}}, \bibinfo {author} {\bibfnamefont {T.}~\bibnamefont {Zhang}}, \bibinfo
  {author} {\bibfnamefont {J.-Y.}\ \bibnamefont {Liu}}, \bibinfo {author}
  {\bibfnamefont {P.-X.}\ \bibnamefont {Chen}}, \ and\ \bibinfo {author}
  {\bibfnamefont {J.-M.}\ \bibnamefont {Yuan}},\ }\href@noop {} {\bibfield
  {journal} {\bibinfo  {journal} {Phys. Rev. Lett.}\ }\textbf {\bibinfo
  {volume} {108}},\ \bibinfo {pages} {170403} (\bibinfo {year}
  {2012})}\BibitemShut {NoStop}%
\bibitem [{\citenamefont {Carrasquilla}\ \emph {et~al.}(2019)\citenamefont
  {Carrasquilla}, \citenamefont {Torlai}, \citenamefont {Melko},\ and\
  \citenamefont {Aolita}}]{carrasquilla2019reconstructing}%
  \BibitemOpen
  \bibfield  {author} {\bibinfo {author} {\bibfnamefont {J.}~\bibnamefont
  {Carrasquilla}}, \bibinfo {author} {\bibfnamefont {G.}~\bibnamefont
  {Torlai}}, \bibinfo {author} {\bibfnamefont {R.~G.}\ \bibnamefont {Melko}}, \
  and\ \bibinfo {author} {\bibfnamefont {L.}~\bibnamefont {Aolita}},\
  }\href@noop {} {\bibfield  {journal} {\bibinfo  {journal} {Nat. Mach.
  Intell.}\ }\textbf {\bibinfo {volume} {1}},\ \bibinfo {pages} {155} (\bibinfo
  {year} {2019})}\BibitemShut {NoStop}%
\bibitem [{\citenamefont {Cha}\ \emph {et~al.}(2020)\citenamefont {Cha},
  \citenamefont {Ginsparg}, \citenamefont {Wu}, \citenamefont {Carrasquilla},
  \citenamefont {McMahon},\ and\ \citenamefont {Kim}}]{cha2020attention}%
  \BibitemOpen
  \bibfield  {author} {\bibinfo {author} {\bibfnamefont {P.}~\bibnamefont
  {Cha}}, \bibinfo {author} {\bibfnamefont {P.}~\bibnamefont {Ginsparg}},
  \bibinfo {author} {\bibfnamefont {F.}~\bibnamefont {Wu}}, \bibinfo {author}
  {\bibfnamefont {J.}~\bibnamefont {Carrasquilla}}, \bibinfo {author}
  {\bibfnamefont {P.~L.}\ \bibnamefont {McMahon}}, \ and\ \bibinfo {author}
  {\bibfnamefont {E.-A.}\ \bibnamefont {Kim}},\ }\href@noop {} {\bibfield
  {journal} {\bibinfo  {journal} {arXiv:2006.12469}\ } (\bibinfo {year}
  {2020})}\BibitemShut {NoStop}%
\bibitem [{\citenamefont {Tiunov}\ \emph {et~al.}(2020)\citenamefont {Tiunov},
  \citenamefont {Tiunova}, \citenamefont {Ulanov}, \citenamefont {Lvovsky},\
  and\ \citenamefont {Fedorov}}]{tiunov2020experimental}%
  \BibitemOpen
  \bibfield  {author} {\bibinfo {author} {\bibfnamefont {E.~S.}\ \bibnamefont
  {Tiunov}}, \bibinfo {author} {\bibfnamefont {V.}~\bibnamefont {Tiunova}},
  \bibinfo {author} {\bibfnamefont {A.~E.}\ \bibnamefont {Ulanov}}, \bibinfo
  {author} {\bibfnamefont {A.}~\bibnamefont {Lvovsky}}, \ and\ \bibinfo
  {author} {\bibfnamefont {A.}~\bibnamefont {Fedorov}},\ }\href@noop {}
  {\bibfield  {journal} {\bibinfo  {journal} {Optica}\ }\textbf {\bibinfo
  {volume} {7}},\ \bibinfo {pages} {448} (\bibinfo {year} {2020})}\BibitemShut
  {NoStop}%
\bibitem [{\citenamefont {Torlai}\ \emph {et~al.}(2019)\citenamefont {Torlai},
  \citenamefont {Timar}, \citenamefont {van Nieuwenburg}, \citenamefont
  {Levine}, \citenamefont {Omran}, \citenamefont {Keesling}, \citenamefont
  {Bernien}, \citenamefont {Greiner}, \citenamefont
  {Vuleti\ifmmode~\acute{c}\else \'{c}\fi{}}, \citenamefont {Lukin},
  \citenamefont {Melko},\ and\ \citenamefont {Endres}}]{torlai2019integrating}%
  \BibitemOpen
  \bibfield  {author} {\bibinfo {author} {\bibfnamefont {G.}~\bibnamefont
  {Torlai}}, \bibinfo {author} {\bibfnamefont {B.}~\bibnamefont {Timar}},
  \bibinfo {author} {\bibfnamefont {E.~P.~L.}\ \bibnamefont {van Nieuwenburg}},
  \bibinfo {author} {\bibfnamefont {H.}~\bibnamefont {Levine}}, \bibinfo
  {author} {\bibfnamefont {A.}~\bibnamefont {Omran}}, \bibinfo {author}
  {\bibfnamefont {A.}~\bibnamefont {Keesling}}, \bibinfo {author}
  {\bibfnamefont {H.}~\bibnamefont {Bernien}}, \bibinfo {author} {\bibfnamefont
  {M.}~\bibnamefont {Greiner}}, \bibinfo {author} {\bibfnamefont
  {V.}~\bibnamefont {Vuleti\ifmmode~\acute{c}\else \'{c}\fi{}}}, \bibinfo
  {author} {\bibfnamefont {M.~D.}\ \bibnamefont {Lukin}}, \bibinfo {author}
  {\bibfnamefont {R.~G.}\ \bibnamefont {Melko}}, \ and\ \bibinfo {author}
  {\bibfnamefont {M.}~\bibnamefont {Endres}},\ }\href {\doibase
  10.1103/PhysRevLett.123.230504} {\bibfield  {journal} {\bibinfo  {journal}
  {Phys. Rev. Lett.}\ }\textbf {\bibinfo {volume} {123}},\ \bibinfo {pages}
  {230504} (\bibinfo {year} {2019})}\BibitemShut {NoStop}%
\bibitem [{\citenamefont {Neugebauer}\ \emph {et~al.}(2020)\citenamefont
  {Neugebauer}, \citenamefont {Fischer}, \citenamefont {J{\"a}ger},
  \citenamefont {Czischek}, \citenamefont {Jochim}, \citenamefont
  {Weidem{\"u}ller},\ and\ \citenamefont
  {G{\"a}rttner}}]{neugebauer2020neural}%
  \BibitemOpen
  \bibfield  {author} {\bibinfo {author} {\bibfnamefont {M.}~\bibnamefont
  {Neugebauer}}, \bibinfo {author} {\bibfnamefont {L.}~\bibnamefont {Fischer}},
  \bibinfo {author} {\bibfnamefont {A.}~\bibnamefont {J{\"a}ger}}, \bibinfo
  {author} {\bibfnamefont {S.}~\bibnamefont {Czischek}}, \bibinfo {author}
  {\bibfnamefont {S.}~\bibnamefont {Jochim}}, \bibinfo {author} {\bibfnamefont
  {M.}~\bibnamefont {Weidem{\"u}ller}}, \ and\ \bibinfo {author} {\bibfnamefont
  {M.}~\bibnamefont {G{\"a}rttner}},\ }\href@noop {} {\bibfield  {journal}
  {\bibinfo  {journal} {Phys. Rev. A}\ }\textbf {\bibinfo {volume} {102}},\
  \bibinfo {pages} {042604} (\bibinfo {year} {2020})}\BibitemShut {NoStop}%
\bibitem [{\citenamefont {Lohani}\ \emph {et~al.}(2020)\citenamefont {Lohani},
  \citenamefont {Kirby}, \citenamefont {Brodsky}, \citenamefont {Danaci},\ and\
  \citenamefont {Glasser}}]{lohani2020machine}%
  \BibitemOpen
  \bibfield  {author} {\bibinfo {author} {\bibfnamefont {S.}~\bibnamefont
  {Lohani}}, \bibinfo {author} {\bibfnamefont {B.~T.}\ \bibnamefont {Kirby}},
  \bibinfo {author} {\bibfnamefont {M.}~\bibnamefont {Brodsky}}, \bibinfo
  {author} {\bibfnamefont {O.}~\bibnamefont {Danaci}}, \ and\ \bibinfo {author}
  {\bibfnamefont {R.~T.}\ \bibnamefont {Glasser}},\ }\href {\doibase
  10.1088/2632-2153/ab9a21} {\bibfield  {journal} {\bibinfo  {journal} {Mach.
  Learn. Sci. Technol.}\ }\textbf {\bibinfo {volume} {1}},\ \bibinfo {pages}
  {035007} (\bibinfo {year} {2020})}\BibitemShut {NoStop}%
\bibitem [{\citenamefont {Xu}\ and\ \citenamefont {Xu}(2018)}]{xu2018neural}%
  \BibitemOpen
  \bibfield  {author} {\bibinfo {author} {\bibfnamefont {Q.}~\bibnamefont
  {Xu}}\ and\ \bibinfo {author} {\bibfnamefont {S.}~\bibnamefont {Xu}},\
  }\href@noop {} {\bibfield  {journal} {\bibinfo  {journal} {arXiv:1811.06654}\
  } (\bibinfo {year} {2018})}\BibitemShut {NoStop}%
\bibitem [{\citenamefont {Lohani}\ \emph {et~al.}(2022)\citenamefont {Lohani},
  \citenamefont {Lukens}, \citenamefont {Glasser}, \citenamefont {Searles},\
  and\ \citenamefont {Kirby}}]{Lohani_2022}%
  \BibitemOpen
  \bibfield  {author} {\bibinfo {author} {\bibfnamefont {S.}~\bibnamefont
  {Lohani}}, \bibinfo {author} {\bibfnamefont {J.~M.}\ \bibnamefont {Lukens}},
  \bibinfo {author} {\bibfnamefont {R.~T.}\ \bibnamefont {Glasser}}, \bibinfo
  {author} {\bibfnamefont {T.~A.}\ \bibnamefont {Searles}}, \ and\ \bibinfo
  {author} {\bibfnamefont {B.~T.}\ \bibnamefont {Kirby}},\ }\href {\doibase
  10.1088/2632-2153/ac9036} {\bibfield  {journal} {\bibinfo  {journal} {Mach.
  Learn. Sci. Technol.}\ }\textbf {\bibinfo {volume} {3}},\ \bibinfo {pages}
  {04LT01} (\bibinfo {year} {2022})}\BibitemShut {NoStop}%
\bibitem [{\citenamefont {Ahmed}\ \emph {et~al.}(2021)\citenamefont {Ahmed},
  \citenamefont {Mu{\~n}oz}, \citenamefont {Nori},\ and\ \citenamefont
  {Kockum}}]{ahmed2021quantum}%
  \BibitemOpen
  \bibfield  {author} {\bibinfo {author} {\bibfnamefont {S.}~\bibnamefont
  {Ahmed}}, \bibinfo {author} {\bibfnamefont {C.~S.}\ \bibnamefont
  {Mu{\~n}oz}}, \bibinfo {author} {\bibfnamefont {F.}~\bibnamefont {Nori}}, \
  and\ \bibinfo {author} {\bibfnamefont {A.~F.}\ \bibnamefont {Kockum}},\
  }\href@noop {} {\bibfield  {journal} {\bibinfo  {journal} {Phys. Rev. Lett.}\
  }\textbf {\bibinfo {volume} {127}},\ \bibinfo {pages} {140502} (\bibinfo
  {year} {2021})}\BibitemShut {NoStop}%
\bibitem [{\citenamefont {Blume-Kohout}(2010)}]{Blume2010}%
  \BibitemOpen
  \bibfield  {author} {\bibinfo {author} {\bibfnamefont {R.}~\bibnamefont
  {Blume-Kohout}},\ }\href {http://stacks.iop.org/1367-2630/12/i=4/a=043034}
  {\bibfield  {journal} {\bibinfo  {journal} {New J. Phys.}\ }\textbf {\bibinfo
  {volume} {12}},\ \bibinfo {pages} {043034} (\bibinfo {year}
  {2010})}\BibitemShut {NoStop}%
\bibitem [{\citenamefont {Seah}\ \emph {et~al.}(2015)\citenamefont {Seah},
  \citenamefont {Shang}, \citenamefont {Ng}, \citenamefont {Nott},\ and\
  \citenamefont {Englert}}]{Seah2015}%
  \BibitemOpen
  \bibfield  {author} {\bibinfo {author} {\bibfnamefont {Y.-L.}\ \bibnamefont
  {Seah}}, \bibinfo {author} {\bibfnamefont {J.}~\bibnamefont {Shang}},
  \bibinfo {author} {\bibfnamefont {H.~K.}\ \bibnamefont {Ng}}, \bibinfo
  {author} {\bibfnamefont {D.~J.}\ \bibnamefont {Nott}}, \ and\ \bibinfo
  {author} {\bibfnamefont {B.-G.}\ \bibnamefont {Englert}},\ }\href {\doibase
  10.1088/1367-2630/17/4/043018} {\bibfield  {journal} {\bibinfo  {journal}
  {New J. Phys.}\ }\textbf {\bibinfo {volume} {17}},\ \bibinfo {pages} {043018}
  (\bibinfo {year} {2015})}\BibitemShut {NoStop}%
\bibitem [{\citenamefont {Granade}\ \emph {et~al.}(2016)\citenamefont
  {Granade}, \citenamefont {Combes},\ and\ \citenamefont {Cory}}]{Granade2016}%
  \BibitemOpen
  \bibfield  {author} {\bibinfo {author} {\bibfnamefont {C.}~\bibnamefont
  {Granade}}, \bibinfo {author} {\bibfnamefont {J.}~\bibnamefont {Combes}}, \
  and\ \bibinfo {author} {\bibfnamefont {D.~G.}\ \bibnamefont {Cory}},\ }\href
  {http://stacks.iop.org/1367-2630/18/i=3/a=033024} {\bibfield  {journal}
  {\bibinfo  {journal} {New J. Phys.}\ }\textbf {\bibinfo {volume} {18}},\
  \bibinfo {pages} {033024} (\bibinfo {year} {2016})}\BibitemShut {NoStop}%
\bibitem [{\citenamefont {Williams}\ and\ \citenamefont
  {Lougovski}(2017)}]{Williams2017}%
  \BibitemOpen
  \bibfield  {author} {\bibinfo {author} {\bibfnamefont {B.~P.}\ \bibnamefont
  {Williams}}\ and\ \bibinfo {author} {\bibfnamefont {P.}~\bibnamefont
  {Lougovski}},\ }\href {http://stacks.iop.org/1367-2630/19/i=4/a=043003}
  {\bibfield  {journal} {\bibinfo  {journal} {New J. Phys.}\ }\textbf {\bibinfo
  {volume} {19}},\ \bibinfo {pages} {043003} (\bibinfo {year}
  {2017})}\BibitemShut {NoStop}%
\bibitem [{\citenamefont {Mai}\ and\ \citenamefont {Alquier}(2017)}]{Mai2017}%
  \BibitemOpen
  \bibfield  {author} {\bibinfo {author} {\bibfnamefont {T.~T.}\ \bibnamefont
  {Mai}}\ and\ \bibinfo {author} {\bibfnamefont {P.}~\bibnamefont {Alquier}},\
  }\href {\doibase https://doi.org/10.1016/j.jspi.2016.11.003} {\bibfield
  {journal} {\bibinfo  {journal} {J. Stat. Plan. Inference}\ }\textbf {\bibinfo
  {volume} {184}},\ \bibinfo {pages} {62} (\bibinfo {year} {2017})}\BibitemShut
  {NoStop}%
\bibitem [{\citenamefont {Lukens}\ \emph {et~al.}(2020)\citenamefont {Lukens},
  \citenamefont {Law}, \citenamefont {Jasra},\ and\ \citenamefont
  {Lougovski}}]{lukens2020practical}%
  \BibitemOpen
  \bibfield  {author} {\bibinfo {author} {\bibfnamefont {J.~M.}\ \bibnamefont
  {Lukens}}, \bibinfo {author} {\bibfnamefont {K.~J.~H.}\ \bibnamefont {Law}},
  \bibinfo {author} {\bibfnamefont {A.}~\bibnamefont {Jasra}}, \ and\ \bibinfo
  {author} {\bibfnamefont {P.}~\bibnamefont {Lougovski}},\ }\href {\doibase
  10.1088/1367-2630/ab8efa} {\bibfield  {journal} {\bibinfo  {journal} {New J.
  Phys.}\ }\textbf {\bibinfo {volume} {22}},\ \bibinfo {pages} {063038}
  (\bibinfo {year} {2020})}\BibitemShut {NoStop}%
\bibitem [{\citenamefont {Lukens}\ \emph {et~al.}(2021)\citenamefont {Lukens},
  \citenamefont {Law},\ and\ \citenamefont {Bennink}}]{lukens2021bayesian}%
  \BibitemOpen
  \bibfield  {author} {\bibinfo {author} {\bibfnamefont {J.~M.}\ \bibnamefont
  {Lukens}}, \bibinfo {author} {\bibfnamefont {K.~J.}\ \bibnamefont {Law}}, \
  and\ \bibinfo {author} {\bibfnamefont {R.~S.}\ \bibnamefont {Bennink}},\
  }\href@noop {} {\bibfield  {journal} {\bibinfo  {journal} {npj Quantum Inf.}\
  }\textbf {\bibinfo {volume} {7}},\ \bibinfo {pages} {113} (\bibinfo {year}
  {2021})}\BibitemShut {NoStop}%
\bibitem [{\citenamefont {Robert}\ and\ \citenamefont
  {Casella}(1999)}]{Robert1999}%
  \BibitemOpen
  \bibfield  {author} {\bibinfo {author} {\bibfnamefont {C.~P.}\ \bibnamefont
  {Robert}}\ and\ \bibinfo {author} {\bibfnamefont {G.}~\bibnamefont
  {Casella}},\ }\href@noop {} {\emph {\bibinfo {title} {Monte Carlo Statistical
  Methods}}}\ (\bibinfo  {publisher} {Springer},\ \bibinfo {address} {New
  York},\ \bibinfo {year} {1999})\BibitemShut {NoStop}%
\bibitem [{\citenamefont {Lohani}\ \emph
  {et~al.}(2021{\natexlab{a}})\citenamefont {Lohani}, \citenamefont {Lukens},
  \citenamefont {Jones}, \citenamefont {Searles}, \citenamefont {Glasser},\
  and\ \citenamefont {Kirby}}]{lohani2021improving}%
  \BibitemOpen
  \bibfield  {author} {\bibinfo {author} {\bibfnamefont {S.}~\bibnamefont
  {Lohani}}, \bibinfo {author} {\bibfnamefont {J.~M.}\ \bibnamefont {Lukens}},
  \bibinfo {author} {\bibfnamefont {D.~E.}\ \bibnamefont {Jones}}, \bibinfo
  {author} {\bibfnamefont {T.~A.}\ \bibnamefont {Searles}}, \bibinfo {author}
  {\bibfnamefont {R.~T.}\ \bibnamefont {Glasser}}, \ and\ \bibinfo {author}
  {\bibfnamefont {B.~T.}\ \bibnamefont {Kirby}},\ }\href@noop {} {\bibfield
  {journal} {\bibinfo  {journal} {Phys. Rev. Research}\ }\textbf {\bibinfo
  {volume} {3}},\ \bibinfo {pages} {043145} (\bibinfo {year}
  {2021}{\natexlab{a}})}\BibitemShut {NoStop}%
\bibitem [{\citenamefont {Cotter}\ \emph {et~al.}(2013)\citenamefont {Cotter},
  \citenamefont {Roberts}, \citenamefont {Stuart},\ and\ \citenamefont
  {White}}]{Cotter2013}%
  \BibitemOpen
  \bibfield  {author} {\bibinfo {author} {\bibfnamefont {S.~L.}\ \bibnamefont
  {Cotter}}, \bibinfo {author} {\bibfnamefont {G.~O.}\ \bibnamefont {Roberts}},
  \bibinfo {author} {\bibfnamefont {A.~M.}\ \bibnamefont {Stuart}}, \ and\
  \bibinfo {author} {\bibfnamefont {D.}~\bibnamefont {White}},\ }\href
  {\doibase 10.1214/13-STS421} {\bibfield  {journal} {\bibinfo  {journal}
  {Statist. Sci.}\ }\textbf {\bibinfo {volume} {28}},\ \bibinfo {pages} {424}
  (\bibinfo {year} {2013})}\BibitemShut {NoStop}%
\bibitem [{\citenamefont {Sommers}\ and\ \citenamefont
  {Zyczkowski}(2003)}]{sommers2003bures}%
  \BibitemOpen
  \bibfield  {author} {\bibinfo {author} {\bibfnamefont {H.-J.}\ \bibnamefont
  {Sommers}}\ and\ \bibinfo {author} {\bibfnamefont {K.}~\bibnamefont
  {Zyczkowski}},\ }\href@noop {} {\bibfield  {journal} {\bibinfo  {journal} {J.
  Phys. A: Math. Gen.}\ }\textbf {\bibinfo {volume} {36}},\ \bibinfo {pages}
  {10083} (\bibinfo {year} {2003})}\BibitemShut {NoStop}%
\bibitem [{\citenamefont {Zyczkowski}\ and\ \citenamefont
  {Sommers}(2003)}]{zyczkowski2003hilbert}%
  \BibitemOpen
  \bibfield  {author} {\bibinfo {author} {\bibfnamefont {K.}~\bibnamefont
  {Zyczkowski}}\ and\ \bibinfo {author} {\bibfnamefont {H.-J.}\ \bibnamefont
  {Sommers}},\ }\href@noop {} {\bibfield  {journal} {\bibinfo  {journal} {J.
  Phys. A: Math. Gen.}\ }\textbf {\bibinfo {volume} {36}},\ \bibinfo {pages}
  {10115} (\bibinfo {year} {2003})}\BibitemShut {NoStop}%
\bibitem [{\citenamefont {Lohani}\ \emph
  {et~al.}(2021{\natexlab{b}})\citenamefont {Lohani}, \citenamefont {Searles},
  \citenamefont {Kirby},\ and\ \citenamefont
  {Glasser}}]{lohani2021experimental}%
  \BibitemOpen
  \bibfield  {author} {\bibinfo {author} {\bibfnamefont {S.}~\bibnamefont
  {Lohani}}, \bibinfo {author} {\bibfnamefont {T.~A.}\ \bibnamefont {Searles}},
  \bibinfo {author} {\bibfnamefont {B.~T.}\ \bibnamefont {Kirby}}, \ and\
  \bibinfo {author} {\bibfnamefont {R.}~\bibnamefont {Glasser}},\ }\href@noop
  {} {\bibfield  {journal} {\bibinfo  {journal} {IEEE Trans. Quantum Eng.}\
  }\textbf {\bibinfo {volume} {2}},\ \bibinfo {pages} {2103410} (\bibinfo
  {year} {2021}{\natexlab{b}})}\BibitemShut {NoStop}%
\bibitem [{\citenamefont {NuCrypt}(2022)}]{nucrypt}%
  \BibitemOpen
  \bibfield  {author} {\bibinfo {author} {\bibnamefont {NuCrypt}},\ }\href
  {http://nucrypt.net/quantum-optical-instrumentation.html} {\enquote {\bibinfo
  {title} {Quantum optical instrumentation},}\ }\bibinfo {howpublished}
  {\url{http://nucrypt.net/quantum-optical-instrumentation.html}} (\bibinfo
  {year} {2022})\BibitemShut {NoStop}%
\bibitem [{\citenamefont {Fiorentino}\ \emph {et~al.}(2002)\citenamefont
  {Fiorentino}, \citenamefont {Voss}, \citenamefont {Sharping},\ and\
  \citenamefont {Kumar}}]{fiorentino2002all}%
  \BibitemOpen
  \bibfield  {author} {\bibinfo {author} {\bibfnamefont {M.}~\bibnamefont
  {Fiorentino}}, \bibinfo {author} {\bibfnamefont {P.~L.}\ \bibnamefont
  {Voss}}, \bibinfo {author} {\bibfnamefont {J.~E.}\ \bibnamefont {Sharping}},
  \ and\ \bibinfo {author} {\bibfnamefont {P.}~\bibnamefont {Kumar}},\
  }\href@noop {} {\bibfield  {journal} {\bibinfo  {journal} {IEEE Photon.
  Technol. Lett.}\ }\textbf {\bibinfo {volume} {14}},\ \bibinfo {pages} {983}
  (\bibinfo {year} {2002})}\BibitemShut {NoStop}%
\bibitem [{\citenamefont {Wang}\ and\ \citenamefont
  {Kanter}(2009)}]{wang2009robust}%
  \BibitemOpen
  \bibfield  {author} {\bibinfo {author} {\bibfnamefont {S.~X.}\ \bibnamefont
  {Wang}}\ and\ \bibinfo {author} {\bibfnamefont {G.~S.}\ \bibnamefont
  {Kanter}},\ }\href@noop {} {\bibfield  {journal} {\bibinfo  {journal} {IEEE
  J. Sel. Top. Quantum Electron.}\ }\textbf {\bibinfo {volume} {15}},\ \bibinfo
  {pages} {1733} (\bibinfo {year} {2009})}\BibitemShut {NoStop}%
\bibitem [{\citenamefont {Banaszek}\ \emph {et~al.}(1999)\citenamefont
  {Banaszek}, \citenamefont {D'Ariano}, \citenamefont {Paris},\ and\
  \citenamefont {Sacchi}}]{Banaszek1999}%
  \BibitemOpen
  \bibfield  {author} {\bibinfo {author} {\bibfnamefont {K.}~\bibnamefont
  {Banaszek}}, \bibinfo {author} {\bibfnamefont {G.~M.}\ \bibnamefont
  {D'Ariano}}, \bibinfo {author} {\bibfnamefont {M.~G.~A.}\ \bibnamefont
  {Paris}}, \ and\ \bibinfo {author} {\bibfnamefont {M.~F.}\ \bibnamefont
  {Sacchi}},\ }\href {\doibase 10.1103/PhysRevA.61.010304} {\bibfield
  {journal} {\bibinfo  {journal} {Phys. Rev. A}\ }\textbf {\bibinfo {volume}
  {61}},\ \bibinfo {pages} {010304} (\bibinfo {year} {1999})}\BibitemShut
  {NoStop}%
\bibitem [{\citenamefont {Altepeter}\ \emph {et~al.}(2005)\citenamefont
  {Altepeter}, \citenamefont {Jeffrey},\ and\ \citenamefont
  {Kwiat}}]{altepeter2005photonic}%
  \BibitemOpen
  \bibfield  {author} {\bibinfo {author} {\bibfnamefont {J.~B.}\ \bibnamefont
  {Altepeter}}, \bibinfo {author} {\bibfnamefont {E.~R.}\ \bibnamefont
  {Jeffrey}}, \ and\ \bibinfo {author} {\bibfnamefont {P.~G.}\ \bibnamefont
  {Kwiat}},\ }\href@noop {} {\bibfield  {journal} {\bibinfo  {journal} {Adv.
  At. Mol. Opt. Phys.}\ }\textbf {\bibinfo {volume} {52}},\ \bibinfo {pages}
  {105} (\bibinfo {year} {2005})}\BibitemShut {NoStop}%
\bibitem [{\citenamefont {Lu}\ \emph {et~al.}(2020)\citenamefont {Lu},
  \citenamefont {Simmerman}, \citenamefont {Lougovski}, \citenamefont
  {Weiner},\ and\ \citenamefont {Lukens}}]{lu2020fully}%
  \BibitemOpen
  \bibfield  {author} {\bibinfo {author} {\bibfnamefont {H.-H.}\ \bibnamefont
  {Lu}}, \bibinfo {author} {\bibfnamefont {E.~M.}\ \bibnamefont {Simmerman}},
  \bibinfo {author} {\bibfnamefont {P.}~\bibnamefont {Lougovski}}, \bibinfo
  {author} {\bibfnamefont {A.~M.}\ \bibnamefont {Weiner}}, \ and\ \bibinfo
  {author} {\bibfnamefont {J.~M.}\ \bibnamefont {Lukens}},\ }\href@noop {}
  {\bibfield  {journal} {\bibinfo  {journal} {Phys. Rev. Lett.}\ }\textbf
  {\bibinfo {volume} {125}},\ \bibinfo {pages} {120503} (\bibinfo {year}
  {2020})}\BibitemShut {NoStop}%
\bibitem [{\citenamefont {Lingaraju}\ \emph {et~al.}(2021)\citenamefont
  {Lingaraju}, \citenamefont {Lu}, \citenamefont {Seshadri}, \citenamefont
  {Leaird}, \citenamefont {Weiner},\ and\ \citenamefont
  {Lukens}}]{lingaraju2021adaptive}%
  \BibitemOpen
  \bibfield  {author} {\bibinfo {author} {\bibfnamefont {N.~B.}\ \bibnamefont
  {Lingaraju}}, \bibinfo {author} {\bibfnamefont {H.-H.}\ \bibnamefont {Lu}},
  \bibinfo {author} {\bibfnamefont {S.}~\bibnamefont {Seshadri}}, \bibinfo
  {author} {\bibfnamefont {D.~E.}\ \bibnamefont {Leaird}}, \bibinfo {author}
  {\bibfnamefont {A.~M.}\ \bibnamefont {Weiner}}, \ and\ \bibinfo {author}
  {\bibfnamefont {J.~M.}\ \bibnamefont {Lukens}},\ }\href@noop {} {\bibfield
  {journal} {\bibinfo  {journal} {Optica}\ }\textbf {\bibinfo {volume} {8}},\
  \bibinfo {pages} {329} (\bibinfo {year} {2021})}\BibitemShut {NoStop}%
\bibitem [{\citenamefont {Alshowkan}\ \emph {et~al.}(2021)\citenamefont
  {Alshowkan}, \citenamefont {Williams}, \citenamefont {Evans}, \citenamefont
  {Rao}, \citenamefont {Simmerman}, \citenamefont {Lu}, \citenamefont
  {Lingaraju}, \citenamefont {Weiner}, \citenamefont {Marvinney}, \citenamefont
  {Pai}, \citenamefont {Lawrie}, \citenamefont {Peters},\ and\ \citenamefont
  {Lukens}}]{alshowkan2021reconfigurable}%
  \BibitemOpen
  \bibfield  {author} {\bibinfo {author} {\bibfnamefont {M.}~\bibnamefont
  {Alshowkan}}, \bibinfo {author} {\bibfnamefont {B.~P.}\ \bibnamefont
  {Williams}}, \bibinfo {author} {\bibfnamefont {P.~G.}\ \bibnamefont {Evans}},
  \bibinfo {author} {\bibfnamefont {N.~S.}\ \bibnamefont {Rao}}, \bibinfo
  {author} {\bibfnamefont {E.~M.}\ \bibnamefont {Simmerman}}, \bibinfo {author}
  {\bibfnamefont {H.-H.}\ \bibnamefont {Lu}}, \bibinfo {author} {\bibfnamefont
  {N.~B.}\ \bibnamefont {Lingaraju}}, \bibinfo {author} {\bibfnamefont {A.~M.}\
  \bibnamefont {Weiner}}, \bibinfo {author} {\bibfnamefont {C.~E.}\
  \bibnamefont {Marvinney}}, \bibinfo {author} {\bibfnamefont {Y.-Y.}\
  \bibnamefont {Pai}}, \bibinfo {author} {\bibfnamefont {B.~J.}\ \bibnamefont
  {Lawrie}}, \bibinfo {author} {\bibfnamefont {N.~A.}\ \bibnamefont {Peters}},
  \ and\ \bibinfo {author} {\bibfnamefont {J.~M.}\ \bibnamefont {Lukens}},\
  }\href {\doibase 10.1103/PRXQuantum.2.040304} {\bibfield  {journal} {\bibinfo
   {journal} {PRX Quantum}\ }\textbf {\bibinfo {volume} {2}},\ \bibinfo {pages}
  {040304} (\bibinfo {year} {2021})}\BibitemShut {NoStop}%
\bibitem [{\citenamefont {Al~Osipov}\ \emph {et~al.}(2010)\citenamefont
  {Al~Osipov}, \citenamefont {Sommers},\ and\ \citenamefont
  {{\.Z}yczkowski}}]{al2010random}%
  \BibitemOpen
  \bibfield  {author} {\bibinfo {author} {\bibfnamefont {V.}~\bibnamefont
  {Al~Osipov}}, \bibinfo {author} {\bibfnamefont {H.-J.}\ \bibnamefont
  {Sommers}}, \ and\ \bibinfo {author} {\bibfnamefont {K.}~\bibnamefont
  {{\.Z}yczkowski}},\ }\href@noop {} {\bibfield  {journal} {\bibinfo  {journal}
  {J. Phys. A: Math. Theor.}\ }\textbf {\bibinfo {volume} {43}},\ \bibinfo
  {pages} {055302} (\bibinfo {year} {2010})}\BibitemShut {NoStop}%
\bibitem [{\citenamefont {Mezzadri}(2007)}]{Mezzadri2007}%
  \BibitemOpen
  \bibfield  {author} {\bibinfo {author} {\bibfnamefont {F.}~\bibnamefont
  {Mezzadri}},\ }\href@noop {} {\bibfield  {journal} {\bibinfo  {journal} {Not.
  Am. Math. Soc.}\ }\textbf {\bibinfo {volume} {54}},\ \bibinfo {pages} {592}
  (\bibinfo {year} {2007})}\BibitemShut {NoStop}%
\end{thebibliography}%

\end{document}